\documentclass[conference]{IEEEtran}
\usepackage[whole]{bxcjkjatype} 
\usepackage[unicode]{hyperref} 

\usepackage{amssymb}
\usepackage[overload]{empheq}

\ifCLASSINFOpdf
\else
\fi
\usepackage{url}

\begin{document}
%
\title{脳血腫の分類: セグメンテーションと分類のジョイント学習}

\author{\IEEEauthorblockN{平野北斗}
\IEEEauthorblockA{九州工業大学}
\and
\IEEEauthorblockN{大北 剛}
\IEEEauthorblockA{九州工業大学}
}


%


\maketitle

\begin{abstract}
脳血腫は6-24時間に急成長を遂げ, 予測を誤り脳外科医が手術しなければ命を
落とす. しかし, 脳血腫には急成長をする型と急成長に至らない型の2通りが
存在し, CTなどによる検査を用いて急成長にいたる脳血腫か否かを人工知能で
分析する技術の開発を行う. この脳血腫の分類問題においては, 正例の少ない
分類問題であること, 血腫の形状が不定であること、その他, 不均衡分類問題,
共変量シフト, 少量データ問題, 疑似相関問題などさまざまなものが存在する.
このため, 単純なVGGなどのCNNを用いた分類においては精度を出すことが難し
い. そこで, 本論文においては, 血腫のセマンティックセグメンテーションと
分類とのジョイント学習を行う方法を提案して, 性能を評価した.

\end{abstract}


%
\IEEEpeerreviewmaketitle

\section{はじめに}

脳血腫は6-24時間に急成長を遂げ, 予測を誤り脳外科医が手術しなければ命を
落とす. しかし, 脳血腫には急成長をする型と急成長に至らない型の2通りが
存在し, 脳血腫か否かはCTなどによる検査を最低限必要となる. 急成長をする
型の脳血腫であることを判定するには通常の医師には難しく, 専門の脳外科医
が判定を行う必要のあるケースが多く存在する. CTは脳血腫が存在するかない
かの簡便な検査方法として広く知られており、しかも、大きい病院だけでなく,
小さな病院においても検査が可能となる. このため, われわれのターゲットと
するシナリオは, 小さな病院においてCTで撮影を行ない, このCT画像を人工知
能で分析し, もし疑わしければ, 大学病院などの脳外科の専門医に患者を運び
込むというシナリオである. 

このシナリオに従って, CTでの撮像画像を与えられ, 急成長する可能性のある
血腫のどのクラスに属するかを判定するタスクを本論文では扱う. 
この急発達の予測は典型的に頻度の少ない事象を学習する困難な問題となるだ
けでなく, 形状が動的な変化をしているかのようにフレキシブルな形状を取る.

本論文の構成は以下の通りである. 第2節では, 血腫の分類の歴史的なコンテ
クストと急成長のための分類法をレビューする. 第3節ではセグメンテーショ
ンと分類のジョイント学習法を提案し, 性能を評価する.
第4節では, 結論と今後の課題について述べる.

\section{関連手法}
\section{血腫分類タスク}

\subsection{血腫分類の歴史的コンテクスト}
  
急性脳内血腫(ICH)は脳小血管病(SVD)が発展したもので, このうち3分の1の患
者においてはさらに出血が持続する. 臨床的にこの脳内血腫の増大の可能性を
初期に検知できることは患者の治療に非常に役立つと考えられている.  脳血
腫の増大の最も単純なモデルにおいては、まず, 初期症状において小血管の一
つが血腫を形成する. 次に, 二次的な機械的な剪断損傷などを原因として出血
が持続して広がる. さらに, ヘマトクリット値と蛋白質により密度が個々のケー
スにより異なる形で凝集する. このとき, 見た目は柔組織の傷害、出血による
脳浮腫、脳室系の拡大などの隣接する構造に依存し, 呈する形状はこれらに依
存してかなり異なる. このモデルからも示唆されるように, 脳血腫の形成は以
下のような特徴をもつ.  1つ目は時系列で血腫拡大が形成されることである.
2つ目は最終的な出血の体積は周辺部の損傷した血管の個数によることである.
3つ目は拡大する方向は軸の向き通りではなく, 拡大した血腫の形状は非定型
となることである.

脳血腫の分類にはさまざまな方式が考えられてきた. 1つ目, マージンの形状
の分類を記述する初期の論文は形状の分類を題材とし, 出血の拡大の影響を直
接の題材としていないが, 後の研究により関係が報告されている. これらは以
下の3つの論文である.
\cite{Fujii94}
は脳血腫の形状による分類を試
みた.  丸形状(丸形状のスムーズなマージン), 非定型形状(非定型形状のマル
チノードのマージン), 分離した形状(定型と非定型を二分するフルイドレベル)の
3つである. \cite{Barras09}は急性脳内血腫(ICH)の形状を定型から非定
型までの5段階で分類した.  \cite{Blacquiere15}は同じスケールを用い
て, 非定型なマージンと出血の拡大は独立な関係にあることを考察した. 2つ
目, 血腫拡大がヘテロな出血から来ることを示唆する最初の率直な証拠は,
\cite{Kim08}のスワールサインの分類による. 3つ目, \cite{Li16}
はスワールサインにさらなる条件を課し, ブラックホールサインを
定義した.  このブラックホールサインは高い特異値において出血拡大に関係
していることを示した.

本論文においては, 出血拡大に関係性が高いと考えら
れるハイポデンシティー(hypodensities), マージンイレギュラーサイン
(margin irregularity), ブレンドサイン(blend sign), フルイドレベル
(fluid levels)の4つのクラスに対して分類を行なう\cite{Boulouis16}.

\subsection{ICH拡張に焦点を絞った血腫分類}

図2はBoulouisら\cite{Boulouis16}の以下の4つのクラスへの分類を示す.
  \begin{itemize}
  \item A: ハイポデンシティー(hypodensities, swirl sign/black hole sign/central hypodensity)
  \item B: マージンイレギュラーサイン(Intracerebral hemorrhage with irregular margins and ICH with heterogeneous densities)
  \item C: ブレンドサイン
    \item D: フルイドレベル
  \end{itemize}
この分類に従って血腫分類タスクを上記4つのクラスに分類するタスクとして
考える.  つまり, 与えられた時刻$t$における患者$h_i$の脳CT画像列に対し
て, ハイポデンシティー, マージンイレギュラーサイン, ブレンドサイン, フ
ルイドレベルの4つのクラスに対して分類を行なうタスクである.
  
血腫をこれら4クラスに分類をするBoulouisらの目的の一つは, これらのそれ
ぞれの血腫の分類の型のICHの拡張率の違いである. つまり, ハイポデンシ
ティー: 20.2\%, マージンイレギュラーサイン: 14.0\%, ブレンドサイン:
35\%, フルイドレベル: (記述なし)という拡張率となるからである.  なお,
これらの4クラスは重複可能である. 血腫は複数のクラスに同時に属するモデ
ルを考え,各々のクラスを独立とみなし二値分類問題と考える. つまりマルチ
クラスの分類ではない.

ハイポデンシティー(図3参照)は出血の中にハイポデンスな構造が存在することを総称す
る. 具体的にはスワールサイン, ブラックホールサイン, ヘテロな密度スケー
ル(density heterogeneity scale)という, いずれも同じ現象を捉えていると
考えられているサインがあるかないかで判断される. スワールサインは飲み込
むような見え方,ブラックホールサインはブラックホールが中心に存在するか
のような見え方,ヘテロな密度スケールはヘテロな出血かホモジニアスな出血
かを１から５までのスケールで評価する指標をさす.

ブレンドサイン\cite{Li16}は血腫の中に, ハイパーアテニュエイトな領域と
隣接するハイパーアテニュエイトな領域が混ざったものをいう. 17\%の患者か
ら, このブレンドサインの存在は出血拡大を示唆するという報告もある.

マージンイレギュラーサインは,出血の形状が丸型の形状や分離した形状では
なく非定型な形状をいう\cite{Fujii94}. このサインは出血が成熟して
いないことを視覚的に捉えながら, 周辺部における血腫の境界における二次的
な出血に影響されている可能性はある.

ハイポデンシティーは球状で, 中心が一つで, その形状で一度だけ拡張して収
縮する. 中心に黒い点(ブラックホール)があれば, 拡張率は高い. イレギュラー
マージンの形状は(複数の中心を持ち)丸ではなく, 若干ひしゃげている. ブレ
ンドサインはハイポデンシティーと類似するが, 形状は若干小さく, 白みがかっ
ている. 拡張率は高い. フルイドレベルは血腫が(黒みがかって)液状化する図
となる.

なお、血腫の形状は実際には3D空間にこれが存在する. このため, 血腫の形状
が球である場合, CT画像で丸であることと同時にCT画像と直行する方向におい
ても, 球程度である必要がある.

また, それぞれ4種類の分類は排他的関係にある. つまり重複を許す.  したがっ
て,目標としては, これらの重複を考慮したクラスを増設した. つまり, CT画
像において, ハイポデンシティーとブレンドサインの両方に当て嵌るクラスで
あれば, これをクラスとして設定した. このやり方で用いて分類すると, ほと
んどのブレンドサインのみ, フルイドレベルのみという分類は非常に少なく,
大半のCT画像は他のクラスとの混合クラス、もしくは, 血腫なしという分類と
なる. ただ、血腫の形状がハイポデンシティーが球状で, マージンイレギュラー
サインが楕円状であるという違いからこれらが同時に起こるというのはあまり
ない.

\subsection{血腫ICH拡張タスク}

ICHが拡大するか否かは, 現実的な問題として非常に重要である. また,脳血腫
にはIVHという指標も関係するが, 本論文ではCT撮影時の体積を測定し, それ
に基づいた時間間隔において, ICHが30\%以上拡大したか否かを判定する. 他
に30ml以上増加したなどが議論されることもあるが, ここではICH30\%以上の
拡大のみを論ずる.

このプロトコルはいくつかの機械学習的な困難さを伴う.  1つ目, この間隔は
実際の患者の撮影条件によるため, まちまちである. これは正確には時間ごと
に変化するはずだが, ここではこれを議論しない. これを突き詰めると欠損値
問題になるが, ここでは単純な教師あり学習として扱う. 2つ目, ICH30\%以上
拡大するか否かという2値分類のみの形で論ずる. 3つ目, 本来, 個体毎にICH
が30\%以上拡張するか否かということを議論しているが, スライスで見た場合
にはあるスライスではICH30\%以上するかもしれないが, 他のスライスでは拡
張する必要はない. 画像のみの情報に基づき拡張性を議論する場合, スライス
に拡張があるかないかはどちらかといえば, ないように思える. 血腫はある部
位で異常が起こったとしても, その部位以外の所で起こる事象とは独立であること
が多いと考えられる. 一方, 血腫がある場合, その血腫が一定時間後に拡張する
かどうかは画像内に表われる可能性がある.

\section{手法}

\subsection{データセットと前処理}

CTスキャンは、12の施設から収集した。各被験者の情報は匿名化した。CTスキャンにおける血腫の範囲や位置、状態などのアノテーションは、12の施設の専門医でそれぞれ別で行われた。

CTスキャンのスライス画像のサイズは、512$\times$512である。CTスキャンのデータは、DICOM形式で保存されており、ピクセル値の単位はハウンズフィールド単位（HU）である。1つ目の前処理として、式\ref{equa:hu}のように、対象のHU範囲を選択してコントラストを調整した。

\begin{equation}
    \label{equa:hu}
    I(i,j) = 
    \begin{cases}
    0 & if I_{HU}(i,j) < a\\
    \frac{I_{HU}(i,j)-a}{b-a} \times 255 & if a \leq I_{HU}(i,j) \leq b\\
    255 & if I_{HU}(i,j) > b
    \end{cases}
\end{equation}

ここで、$I(x)$は、位置$x$でのコントラスト調整後の強度である。$a=0$と$b=80$は、脳のCT画像を可視化するために一般的に使用される。本研究では、CTスキャンを収集したそれぞれの施設の専門医が決定したパラメータ$a$,$b$を用いており、おおよそ$a=0$と$b=80$であった。

さらに、CT画像の前処理の記事（\url{https://vincentblog.xyz/posts/medical-images-in-python-computed-tomography}）に従って、ノイズ除去、骨除去、中央への位置合わせの前処理を行った。ノイズ除去では、CTスキャンの際に映るアーイファクトを除去した。骨除去では、頭蓋骨部分を除去した。これは、コントラスト調整後のCT画像において骨の部分が白く映り、情報が強くなる（白は最大のピクセル値255）と考えたためである。中央への位置合わせでは、脳を画像の中央に平行移動する作業を行った。

\subsection{セグメンテーションと分類のためのジョイントモデル}

\begin{figure}[h]
\includegraphics[width=60mm]{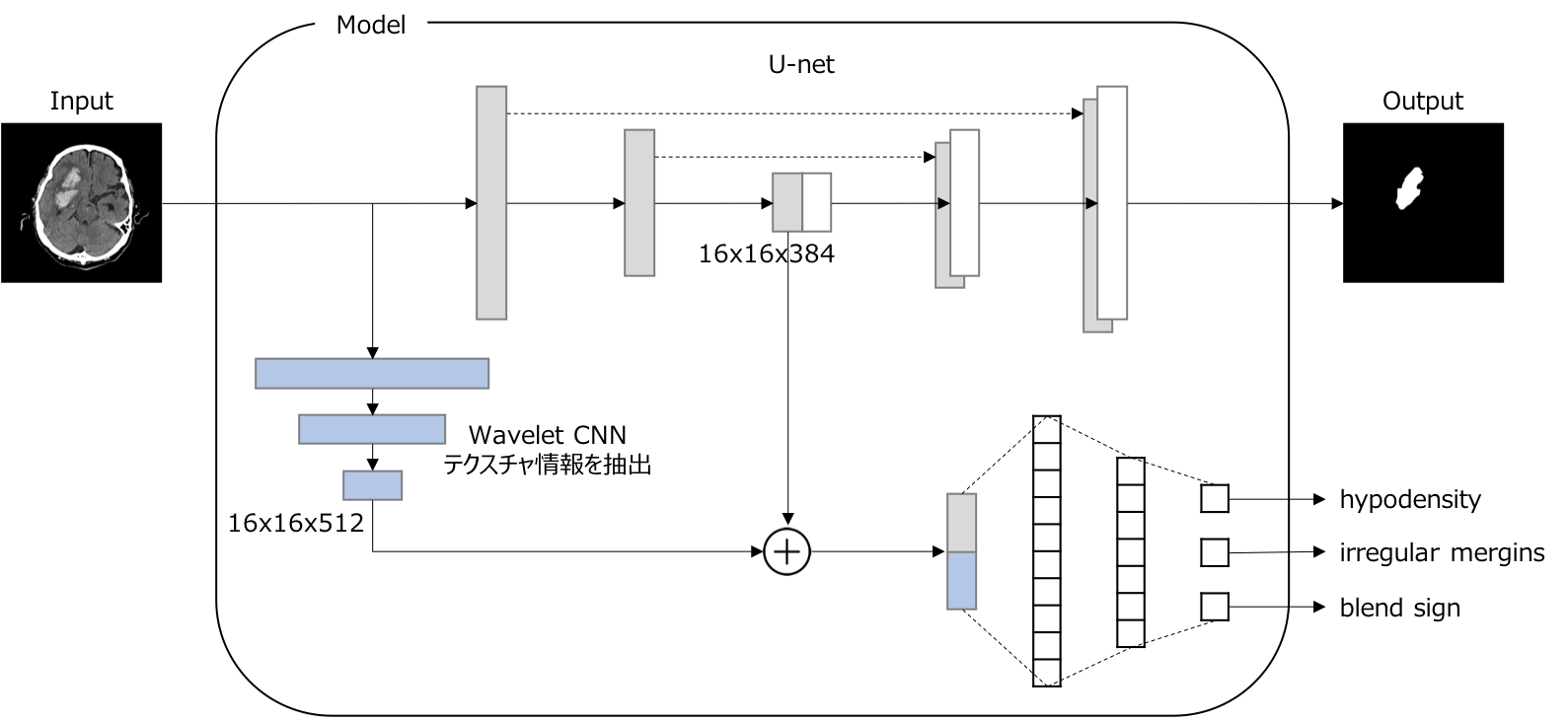}   
\caption{
提案するセグメンテーションと分類のためのジョイントモデルのアーキテクチャ
}
\label{fig:model}
\end{figure}

図\ref{fig:model}は、本研究で提案するセグメンテーションと分類のためのジョイントモデルのアーキテクチャを示している。提案するモデルは、1つのモデルでセグメンテーションと分類の2つのタスクを行う。具体的に、U-Net（図\ref{fig:model}の上のパス）でセグメンテーションのタスクを行い、Wavelet CNN（図\ref{fig:model}の下のパス）で分類のタスクを行う。ここで分類のタスクでは、U-Netのエンコーダより得られる特徴と、Wavelet CNNより得られる特徴の2つを用いて分類を行う。提案するモデルへの入力は前処理後の2DのCT画像であり、セグメンテーションタスクの出力は確率マップ（[0,1]）、分類タスクの出力は確率ベクトル（[0,1]）である。ここで、確率マップの各位置の値は対応するピクセルが属する確率を示す（血腫領域の場合は1、それ以外は0）。確率ベクトルの各要素の値は、対応する血腫状態である確率を示す。

\vspace{1cm}

\subsubsection*{U-Net}
血腫領域のセグメンテーションタスクには、エンコーダ/デーコーダ構造のCNNを採用している。ダウンサンプリングのエンコーダには、Efficientnet-b3モデルを用いている。また、U-Netに従い、エンコーダとデーコーダの対応するレイヤ間にスキップ接続を追加している。

\subsubsection*{Wavelet CNN}
血腫にはテスチャにも情報を持っている。そのため、血腫状態の分類タスクには、一般的なCNNではなく、Wavelet CNNを採用している。Wavelet CNNの中心的なアイデアは、ウェーブレット変換をCNNアーキテクチャに組み込んで、画像のテクスチャ情報を保持することである。一般的なCNNでは、受容野を拡大するために用いられるプーリング処理の際に、エッジ情報が保持され、テクスチャ情報は損失する可能性がある。本研究では、VGG16モデルのプーリング層をウェーブレット変換に変更したCNNをWavelet CNNとして扱う。

\subsubsection*{マルチラベルの分類}
血腫状態の分類タスクは、各CT画像が1つずつのクラスに属する多クラス分類ではなく、いくつかのクラスに属するマルチラベルの分類タスクである。一般的な多クラス分類では、出力層の活性化関数としてsoftmax関数を用いる。一方で、マルチラベルの分類タスクではsigmoid関数を用いる。これは明確で、softmax関数の定義が、sigmoidの和で割ったものだからである。softmax関数の出力$y_i$の合計は必ず1になるが、sigmoid関数の出力$y_i$の合計が1になる保証はない。sigmoid関数の出力$y_i$は、$y_i$のそれぞれが[0,1]の範囲の値になる。

また、一般的な多クラス分類では損失関数として多クラス交差エントロピーを用いることが多い。一方で、マルチラベルの分類では二値交差エントロピーを用いる。本研究では、血腫状態のラベル分布が不均衡なため、式\ref{equa:loss}のような重み付きの二値交差エントロピーを用いる。

\begin{equation}
    \begin{split}
    \label{equa:loss}
    & L(x,y) = \\ 
    & - \frac{1}{NC} \sum_{n=1}^{N} \sum_{c=1}^{C} [w_c y_{n,c} \cdot log(x_{n,c}) \\
    & + (1-y_{n,c}) \cdot log(1-x_{n,c})] \\
    & weight(c) = \frac{sum(negative examples(c))}{sum(positive examples(c))}
    \end{split}    
\end{equation}

\subsubsection*{学習手順}
提案するモデルは、2段階の学習手順を採用している。第1段階では、セグメンテーションを行うU-Netを学習する。第2段階では、分類を行うWavelet CNNを学習する。この際、U-Netのエンコーダより得られる特徴も使用するが、U-Netの重みは固定している。

\section{実験}

\subsection{血腫領域のセグメンテーション}
\subsubsection{データセットの分割方法や前処理の影響}

データセットの分割方法や前処理の影響を評価した。CT画像を訓練用とテスト用に分割する方法として、ランダムに分割する場合と、患者ごとに分割する場合の2つを考えた。
また、CT画像の前処理がセグメンテーションの精度に対してどのような影響を与えるのかを評価した。

データセットを患者ごとに分割する場合よりも、ランダムに分割した場合の方が精度が低かった（表x）。この結果は、ランダムに分割した場合には、訓練データセットとテストデータセット間で似た画像が含まれる可能性があるためである。つまり、ランダムに分割すると同じ患者から得られる複数枚のスライス画像が、訓練データセットとテストデータセットに含まれる可能性がある。また、CT画像の前処理を行うことで、精度が向上することを確認した（表x）。

\begin{table}[h]
\begin{tabular}{|l|l|l|l|}
\hline
                                                                     & dice loss & iou    & fscore \\ \hline
ランダム分割                                                               & 0.1367    & 0.7809 & 0.8633 \\ \hline
患者分割                                                                 & 0.1609    & 0.7573 & 0.8390 \\ \hline
\begin{tabular}[c]{@{}l@{}}患者分割\\ + ノイズ除去\\ + 骨除去\\ + 中央への位置合わせ\end{tabular} & 0.1464    & 0.7660 & 0.8536 \\ \hline
\end{tabular}
\end{table}

\subsubsection{血腫位置を考慮した学習}

血腫の位置を考慮してモデルを学習させることで、精度が向上するのかを評価した。具体的には、まず初めに、血腫の位置を考慮せずにモデルの学習を行い学習済みモデルを作成する。その後、血腫の位置xのデータセットのみを用いて、学習済みモデルをファインチューニングさせる。データセットは患者ごとで分割を行い、前処理は行なっていない。

血腫の位置に従ってファインチューニングを行なったが、精度の向上は見られなかった。具体的には、血腫の位置を考慮せずに学習を行なったモデルの、「位置a. 被殻」のデータセットに対する精度は0.9094であり、「位置a. 被殻」のデータセットのみを用いてファインチューニングを行なったが、精度は0.88904と低下した。同様に他の血腫の位置についても、ファインチューニングを行うことによる精度の向上は見られなかった。

\begin{table}[h]
\begin{tabular}{|l|l|l|l|}
\hline
条件                                                                & \begin{tabular}[c]{@{}l@{}}位置a.\\ 被殻\end{tabular} & \begin{tabular}[c]{@{}l@{}}位置b.\\ 視床\end{tabular} & \begin{tabular}[c]{@{}l@{}}位置c.\\ 皮質下\end{tabular} \\ \hline
全データで学習                                                           & 0.9094                                            & 0.8524                                            & 0.8726                                             \\ \hline
\begin{tabular}[c]{@{}l@{}}位置a. 被殻に従って\\ ファインチューニング\end{tabular}  & 0.8904                                            &                                                   &                                                    \\ \hline
\begin{tabular}[c]{@{}l@{}}位置b. 視床に従って\\ ファインチューニング\end{tabular}  &                                                   & 0.8364                                            &                                                    \\ \hline
\begin{tabular}[c]{@{}l@{}}位置c. 皮質下に従って\\ ファインチューニング\end{tabular} &                                                   &                                                   & 0.8607                                             \\ \hline
\end{tabular}
\end{table}

\subsection{血腫状態の分類}

\subsubsection{Wavelet CNNの影響}
Wavelet CNNによって分類の精度が向上するかを評価した。データセットは12つの施設から得られたCT画像を用い、患者ごとに訓練用とテスト用に分割した。データセットの前処理は行っていない。ジョイントモデルの条件として次の4つを考慮した；1. VGG16、2. 重み付き損失のVGG16、3. Wavelet VGG16、4. 重み付き損失のWavelet VGG16。結果として、重み付きwavelet vgg16の場合、特にhypodensityやirregularで高いaucが得られた（表xxx）。

\begin{table}[h]
\begin{tabular}{|l|l|l|l|}
\hline
条件                & hypodensity & irregular & blend \\ \hline
vgg16             & 0.692       & 0.337     & 0.524 \\ \hline
重み付きvgg16         & 0.699       & 0.471     & 0.545 \\ \hline
wavelet vgg16     & 0.650       & 0.608     & 0.613 \\ \hline
重み付きwavelet vgg16 & 0.702       & 0.684     & 0.576 \\ \hline
\end{tabular}
\end{table}

\subsubsection{血腫領域に絞った血腫状態の分類}
本研究では血腫領域のセグメンテーションを行っている。そこで、血腫状態の分類タスクにおいてWavelet CNNの入力の際に、元のCT画像の血腫領域のみを与えることで、血腫状態の分類精度が向上するかを評価した。具体的に、Wavelet CNNの学習の際は、元のCT画像に血腫領域の正解マスクをかけて得られた血腫領域のみの画像を使用する。テスト時は、元のCT画像に、セグメンテーションモデルの出力マスクをかけて得られた画像を使用する。データセットは4つの施設から得られたCT画像を用いた。結果として、マスクをかけて血腫領域を絞ることで、irregularの分類精度は下がったが、hypodensityの分類精度は向上した（表x）。

\begin{table}[h]
\begin{tabular}{|l|l|l|l|}
\hline
条件                                                                & hypodensity & irregular & blend \\ \hline
重み付きwavelet vgg16                                                 & 0.655       & 0.790     & 0.559 \\ \hline
\begin{tabular}[c]{@{}l@{}}重み付きwavelet vgg16\\ + マスク\end{tabular} & 0.676       & 0.732     & 0.559 \\ \hline
\end{tabular}
\end{table}

\subsubsection{Grad Camの一例}
Wavelet CNNモデルより計算されるGrad Camを示す。Grad Camでは、モデルが見ているであろう箇所が強調される。

\begin{figure}[h]
\begin{center}
\includegraphics[width=80mm]{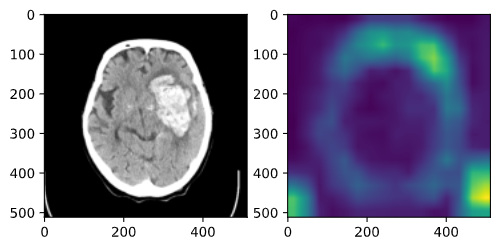} 
\caption{
重み付きwavelet vgg16
}
\end{center}
\end{figure}

\begin{figure}[h]
\begin{center}
\includegraphics[width=80mm]{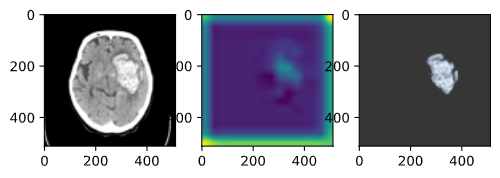} 
\caption{
重み付きwavelet vgg16 + マスク
}
\end{center}
\end{figure}

\section{おわりに}
脳血腫は6-24時間に急成長を遂げ, 予測を誤り脳外科医が手術しなければ命を
落とす. しかし, 脳血腫には急成長をする型と急成長に至らない型の2通りが
存在し, CTなどによる検査を用いて急成長にいたる脳血腫か否かを人工知能で
分析する技術の開発を行う. この脳血腫の分類問題においては, 正例の少ない
分類問題であること, 血腫の形状が不定であること、その他, 不均衡分類問題,
共変量シフト, 少量データ問題, 疑似相関問題などさまざまなものが存在する.
このため, 単純なVGGなどのCNNを用いた分類においては精度を出すことが難し
い. そこで, 本論文においては, 血腫のセマンティックセグメンテーションと
分類とのジョイント学習を行う方法を提案して, 性能を評価した.


\begin{thebibliography}{9}
\bibitem{Blacquiere15}
Blacquiere D, Demchuk AM, Al-Hazzaa M, Deshpande A, Petrcich W, Aviv RI, Rodriguez-Luna D, Molina CA, Silva Blas Y, Dzialowski I, Czlonkowska A, Boulanger JM, Lum C, Gubitz G, Padma V, Roy J, Kase CS, Bhatia R, Hill MD, Dowlatshahi D; PREDICT/Sunnybrook ICH CTA Study Group. Intracerebral Hematoma Morphologic Appearance on Noncontrast Computed Tomography Predicts Significant Hematoma Expansion. Stroke. 2015 Nov;46(11):3111-6. doi: 10.1161/STROKEAHA.115.010566. Epub 2015 Oct 8. PMID: 26451019
  
\bibitem{Boulouis16}
  Boulouis G, et al., Association between hypodensities detected by computed tomography and hematoma expansion in patients with intracerebral hemorrhage. JAMA Neurol 2016;73:961-968. 2016.

\bibitem{Barras09}
  Barras CD, Tress BM, Christensen S, MacGregor L, Collins M, Desmond PM, Skolnick BE, Mayer SA, Broderick JP, Diringer MN, Steiner T, Davis SM; Recombinant Activated Factor VII Intracerebral Hemorrhage Trial Investigators. Density and shape as CT predictors of intracerebral hemorrhage growth. Stroke. 2009 Apr;40(4):1325-31. doi: 10.1161/STROKEAHA.108.536888. Epub 2009 Mar 12. PMID: 19286590.

\bibitem{Li16}
Li Q, Zhang G, Xiong X, Wang XC, Yang WS, Li KW, Wei X, Xie P. Black Hole Sign: Novel Imaging Marker That Predicts Hematoma Growth in Patients With Intracerebral Hemorrhage. Stroke. 2016 Jul;47(7):1777-81. doi: 10.1161/STROKEAHA.116.013186. Epub 2016 May 12. PMID: 27174523

\bibitem{Kim08}
J. Kim, A. Smith, J.C. Hemphill, W.S. Smith, Y. Lu, W.P. Dillon, M. Wintermark.
Contrast Extravasation on CT Predicts Mortality in Primary Intracerebral Hemorrhage. American Journal of Neuroradiology Mar 2008, 29 (3) 520-525; DOI: 10.3174/ajnr.A0859

          



  \bibitem{unet} Ronneberger O., Fischer P., Brox T. (2015) U-Net: Convolutional Networks for Biomedical Image Segmentation. In: Navab N., Hornegger J., Wells W., Frangi A. (eds) Medical Image Computing and Computer-Assisted Intervention – MICCAI 2015. MICCAI 2015. Lecture Notes in Computer Science, vol 9351. Springer, Cham

  \bibitem{fcn} Darrell, JLaESaT, J. Long, and E. Shelhamer. "Fully Convolutional Networks for Semantic Segmentation." IEEE T PATTERN ANAL 39.4 (2014).


  














\bibitem{Fujii94}
Fujii Y, Tanaka R, Takeuchi S, Koike T, Minakawa T, Sasaki O. Hematoma enlargement in spontaneous intracerebral hemorrhage. J Neurosurg. 1994 Jan;80(1):51-7. doi: 10.3171/jns.1994.80.1.0051. PMID: 8271022

\end{thebibliography}
\end{document}